\documentclass{aastex63}
\turnoffedit

\usepackage{CJKutf8}
\received{}
\revised{}
\accepted{}
\submitjournal{AJ}

\shorttitle{DEEP Paper I: Survey description}
\shortauthors{Trilling et al.}
\graphicspath{{./}{figures/}}
\begin{document}

\title{The DECam Ecliptic Exploration Project (DEEP): I. Survey description, science questions, and technical demonstration}

\def \NAU {Department of Astronomy and Planetary Science, Northern Arizona University,\\PO Box 6010, Flagstaff, AZ 86011, USA}
\def \UMPhysics {Department of Physics, University of Michigan,\\ Ann Arbor, MI 48109, USA}
\def \UMAstronomy {Department of Astronomy, University of Michigan,\\ Ann Arbor, MI 48109, USA}
\def \UW {DiRAC Institute and the Department of Astronomy, University of Washington, Seattle, USA}
\def \uchile {Departamento de Astronomía, Universidad de Chile,\\ Camino del Observatorio 1515, Las Condes, Santiago, Chile}
\def \cfa {Harvard-Smithsonian Center for Astrophysics,\\ 60 Garden St., MS 51, Cambridge, MA 02138, USA}
\def \byu {Department of Physics and Astronomy, Brigham Young University, Provo, UT 84602, USA}
\def \apl {Applied Physics Lab, Johns Hopkins University,\\ 11100 Johns Hopkins Road, Laurel, Maryland 20723, USA}
\def \ucla {Department of Earth, Planetary and Space Sciences, University of California Los Angeles, 595 Charles E. Young Dr. East, Los Angeles, CA 90095, USA}
\def \carnegie {Earth and Planets Laboratory, Carnegie Institution for Science, Washington, DC 20015}
\def \stgallen {School of Computer Science, University of St. Gallen,\\ Rosenbergstrasse 30, CH-9000 St. Gallen, Switzerland}

\correspondingauthor{DET}
\email{david.trilling@nau.edu}

\author[0000-0003-4580-3790]{David E. Trilling}
\affiliation{\NAU}

\author[0000-0001-6942-2736]{David~W.~Gerdes}
\affiliation{\UMPhysics}
\affiliation{\UMAstronomy}

\author[0000-0003-1996-9252]{Mario Juri\'c}
\affiliation{\UW}

\author[0000-0001-9859-0894]{Chadwick A. Trujillo}
\affiliation{\NAU}

\author[0000-0003-0743-9422]{Pedro H. Bernardinelli}
\affiliation{\UW}

\author[0000-0003-4827-5049]{Kevin J. Napier}
\affiliation{\UMPhysics}

\author[0000-0002-7895-4344]{Hayden Smotherman}
\affiliation{\UW}

\author[0000-0001-6350-807X]{Ryder Strauss}
\affiliation{\NAU}

\author[0000-0002-5211-0020]{Cesar Fuentes}
\affiliation{\uchile}

\author[0000-0001-8550-6788]{Matthew J. Holman}
\affiliation{\cfa}

\author[0000-0001-7737-6784]{Hsing~Wen~Lin (\begin{CJK*}{UTF8}{gbsn}
林省文\end{CJK*})}
\affiliation{\UMPhysics}

\author[0000-0002-2486-1118]{Larissa Markwardt}
\affiliation{\UMPhysics}

\author{Andrew McNeill}
\affiliation{\NAU}
\affiliation{Department of Physics, Lehigh University, 16 Memorial Drive East, Bethlehem, PA, 18015, USA}

\author[0000-0002-7817-3388]{Michael Mommert}
\affiliation{\stgallen}

\author[0000-0001-5750-4953]{William J. Oldroyd}
\affiliation{\NAU}

\author[0000-0001-5133-6303]{Matthew J. Payne}
\affiliation{\cfa}

\author[0000-0003-1080-9770]{Darin Ragozzine}
\affiliation{\byu}

\author[0000-0002-9939-9976]{Andrew S. Rivkin}
\affiliation{\apl}

\author{Hilke Schlichting}
\affiliation{\ucla}

\author[0000-0003-3145-8682]{Scott S. Sheppard}
\affiliation{\carnegie}

\author[0000-0002-8167-1767]{Fred C.~Adams}
\affiliation{\UMPhysics}
\affiliation{\UMAstronomy}

\author[0000-0001-7335-1715]{Colin Orion Chandler}
\affiliation{\UW}
\affiliation{LSST Interdisciplinary Network for Collaboration and Computing, 933 N. Cherry Avenue, Tucson AZ 85721}
\affiliation{\NAU}



\begin{abstract}
We present here the DECam Ecliptic Exploration Project (DEEP), a three year NOAO/NOIRLab Survey that was allocated 46.5~nights to discover and measure the properties of thousands of trans-Neptunian objects (TNOs) to magnitudes as faint as VR$\sim$27, corresponding to sizes as small as 20~km diameter.
In this paper we present the science goals of this project, the experimental design of our survey, and a technical demonstration of our approach.
The core of our project is ``digital tracking,'' in which all collected images are combined at a range of motion vectors to detect unknown TNOs that are fainter than the single exposure depth of VR$\sim$23~\edit1{mag}.
Through this approach we reach a depth that is approximately 2.5~magnitudes fainter than the standard LSST ``wide fast deep'' nominal survey depth of~24.5~\edit1{mag}.
DEEP will more than double the number of known TNOs with observational arcs of 24~hours or more, and increase by a factor of~10 or more the number of known small ($<$50~km) TNOs. We also describe our ancillary science goals, including 
measuring the mean shape distribution of very small main belt asteroids,
and briefly outline a set of forthcoming papers that present further aspects of and preliminary results from the DEEP program.
\end{abstract}



\section{Introduction}

\noindent The space beyond the orbit of Neptune is inhabited by planetesimals left over from our Solar System's formation 4.5~billion years ago. These trans-Neptunian objects (TNOs --- sometimes also called Kuiper Belt Objects, or KBOs) preserve many clues 
about the evolution of our planetary system.
TNOs shine in reflected light, which makes observing small and distant objects very challenging. As a result, our knowledge of the trans-Neptunian region is based predominantly on the study of only the brightest objects, with magnitudes brighter than around 23--24 and diameters greater than 100--300 km \citep{brown2012}. Our understanding of the smallest (and most numerous) TNOs is generally driven by theoretical models \citep{schlichting2013}, indirect evidence
\citep{schlichting2012,chang2013,liu2015}, and analysis of the sparse existing data
\citep{shankman2013,belton2014}.

\citet{B04} established the state of the art in faint TNO searches by surveying 0.02 deg$^2$ with the Hubble Space Telescope (HST) ACS/WFC camera,
using more than 100 orbits to discover three new objects to a limiting magnitude of R=28.5. This result implied a break in the size distribution of TNOs and hinted at a different size distribution for high 
($i$$>$5$^\circ$) and low ($i$$<$5$^\circ$) inclination objects.
Critically, Bernstein used a shift-and-stack technique to detect TNOs fainter than the noise threshold in individual images.
\citet{fuentes2010,fuentes2011}
placed further constraints on the small TNO population by using archival HST data to detect objects smaller than the break in the size distribution. 
\citet{fraser2014}
compiled data from multiple surveys to refine the absolute magnitude distribution of TNOs, and
\citet{parker2015}
used the New Horizons flyby target search data to update the measurement of the small TNO size distribution. That measurement in turn was used for the assumed impactor flux to constrain the geophysics of Pluto based on its distribution of craters
\citep{trilling2016,mckinnon2016}.

The primordial nature of trans-Neptunian space provides an opportunity to understand the origin and evolution of the Solar System through detailed studies of TNOs
\citep[e.g.,][]{morbidelli2020}.
Some outstanding questions about trans-Neptunian space include the following:
(1) To what extent did the giant planets migrate during the formation of the Solar System? (2) What does the TNO binary fraction imply about the formation of the outer Solar System?
(3) What are the dynamical pathways from the outer Solar System to the inner Solar System? 
(4) What is the origin of the observed range (and dichotomy) of colors in the outer Solar System
\citep[e.g.,][]{doressoundiram2008},
and what does this imply about the formation of the Solar System?
These questions highlight
key gaps in our understanding of the formation and evolution of the Solar System that
can be informed and addressed by expanding on previous efforts to observe faint TNOs.

The size
distribution of the TNOs can be divided into large objects and small objects with a break or change in the slope of the distribution occurring between these two
populations at roughly 50~km in diameter
(see \citet{B04} and many subsequent references).
The shape of the TNO size distribution has often been attributed to collisions, with objects smaller than the break having disruption timescales shorter than the age of the Solar System
\citep{pansari,Kenyon2004}.
However, some more recent modeling efforts have implied that the break in the TNO size distribution is not a collisional product, but a result of formation through gravitational collapse of a pebble cloud
\citep{nesvorny2019,Robinson2020}.
A measurement of the size distribution of TNOs will provide a clear test of these models, as described below.
%Furthermore, testing 
%models of Solar System formation also has important implications
%for the formation of planetary systems in general.
\edit1{Furthermore, understanding how our Solar System formed may provide insight into the general process of planetary system formation \citep[e.g.,][]{2020tnss.book..351W}.}

The details of the TNO size distribution record the formation and evolution of the Solar System, and differences across dynamical subclasses reveal nuances of that history. Furthermore, knowledge of the compositions and shape distributions of TNOs likewise would provide powerful constraints on the history of the outer Solar System. However, to date there is no large-scale catalog of those properties for TNOs.
Unraveling these many details of the history of our Solar System can be addressed through a single large-area, deep survey of the outer Solar System.

We present here the DECam Ecliptic Exploration Project (DEEP), 
a NOIRLab (formerly NOAO) Survey program that is designed to address many of these science questions.
We were allocated 46.5~nights with 
the Dark Energy Camera (DECam) --- a 3~deg$^2$ imager on the 4-meter Blanco telescope at Cerro Tololo Inter-American Observatory (CTIO) in Chile --- to be executed in the six semesters from 2019A through 2021B (extended through 2023A).
Here we present the details of this new survey as well as some preliminary results, as the first in a series of papers.

In Section~2 we present the overall science goals of our project. 
In Section~3 we summarize our experimental design, and in Section~4 we present our general approach to our data processing; both of these topics are explained more completely in separate papers. Section~5 presents
a technical demonstration of our approach. Section~6 describes our expected survey results, and in Section~7 we advertise a set of forthcoming papers that present various DEEP results in detail.

\section{Science goals}

Our overall science goal is to measure the properties of distant Solar System objects
to increase our understanding of the formation and evolution of the Solar System. Our rich
survey data will provide a number of different kinds of constraints.
Our primary goals of DEEP are the following: (1) to measure the size distribution of faint TNOs down to $\sim$20~km;
(2) to derive the shape distribution of TNOs; and 
(3) to measure these physical properties as a function of dynamical class and of size.
Initially, we had an additional science goal of measuring colors
%(and therefore compositions)
of
TNOs, 
\edit1{with the goal of enabling taxonomic classifications as a function of dynamical state and size,}
but poor weather in Year~1 of the survey forced us to eliminate observations in filters other than VR.
We have an ongoing project to use the Magellan telescopes, together with our DEEP data, to partially address this color question, with results to be presented in a future paper (Strauss et al., in prep.).

Our measured size distribution of TNOs (science goal~1) will allow us to test current theories of the formation of the outer Solar System. One current model \citep{fraser2014} broadly matches expectations of standard hierarchical accretion models like that of \citet{Kenyon2004} in which the size distribution in the $1\lesssim D \lesssim 100$~km size range exhibits a shallow power-law produced by heavy collisional evolution, with fragments piling up below $D\lesssim 1$~km. 
A second model modifies the first to include expectations from
\citet{schlichting2013}, in which an alternate hierarchical accretion route occurs, starting from a significant overabundance of small bodies with diameters $D\sim 1$~km. This results in more vigorous collisional evolution, and an extremely steep size distribution slope at $10\lesssim D \lesssim$20~km, as fragments pile up at small sizes.
Finally, a third model is an alteration of the nominal first model to reflect expectations from measurements of Jupiter Family Comets (JFCs), many of which suggest the possible existence of a further shallowing of the size distribution below $D\sim20$~km \citep{Licandro2016}.

\edit1{We will measure the mean (probabilistic) TNO shape distribution (science goal~2) from a large number of partial lightcurves. 
This will be useful in constraining the collisional history of the trans-Neptunian region.}
%The TNO shape distribution that we will measure (science goal~2) will allow us to constrain the collisional history of the trans-Neptunian region.
%$and potentially the mean density of TNOs.
Measuring these properties, and others, as a function of dynamical class (science goal~3) will allow us to understand details of the chronology and evolution of the outer Solar System.

There are a number of secondary science goals that can be met with the dataset that we will acquire. 
(4) We may be able to measure the size distribution of Centaurs down to around 10~km. Since Centaurs originated in trans-Neptunian space, this size distribution will allow us to extend our measured TNO size distribution to even smaller sizes. However, our survey cadence is not optimized for Centaur detection, so our completeness (and, in particular, linking and orbit determination) will be significantly reduced.
(5) We will measure the shape distribution of thousands of main belt asteroids (as a function of size) through modeling the mean underlying shape that best produces the observed distribution of (partial) lightcurves. This will allow us to test theories about the collisional history of the main asteroid belt.
There is also an outstanding question about the origin of the mismatch between the mean asteroid shape for main belt asteroids larger than 1~km and near Earth asteroids smaller than around 500~m \citep{mcneill2019}. Our survey will produce (partial) lightcurves for main belt asteroids as small as 500~m, which will allow us to directly probe whether the mean shape difference is because of size or collisional environment.
(6) We are likely to identify individual objects of interest
in our survey. This may include objects with unusual orbits or physical properties, and may warrant dedicated follow-up observations, to be carried out separately.
(7) Data from this survey could
be searched for signs of activity, particularly among the Centaur and asteroid samples (objects closer to the Sun than TNOs).
(8) Our survey will produce timeseries photometry for millions of stars
that 
could be searched for exoplanet transits, variable stars, and other stellar astrophysics events.
(9) Our archival deep image pointings could be searched for supernovae or other transient events. All DEEP data is/will be publicly accessible through
AstroArchive\footnote{\url{https://astroarchive.noirlab.edu/}}, the NOIRLab Data Archive, under proposal ID 2019A-0337.

\section{Experimental design}

Our primary science goals dictate the requirements for our experimental design. In summary,
we must observe a large area on the sky, and each visit must include a long continuous stare so that we can apply our digital tracking algorithms and reach the necessary combined image depth.
Complete details of our observing strategy are presented in Trujillo et al.\ (in prep.), which is DEEP Paper II.

\subsection{Required depth and expected
sensitivity}

Our science goals (Section~2) require that we observe a large number of TNOs \edit1{with diameters smaller than $\sim$20~km.}
%with diameters of $\sim$20~km or smaller. 
We used the DECam Exposure Time Calculator (ETC) v7 spreadsheet\footnote{\url{https://noirlab.edu/science/documents/scidoc0493}}
and our experience using DECam to observe Solar System targets to
define 
an observational program that meets this science goal. The DECam ETC spreadsheet reports AB magnitudes; we have converted these to Vega magnitudes for all results and tables given here.
Furthermore, the VR filter (not listed in the ETC) is 76\% wider and has 8\% higher throughput than $g$ (a fiducial comparison filter), which implies a detection threshold that is 0.7~mag fainter. 
This corresponds with our experience using these filters with DECam.
At a distance of 40~au, 
a 24~km diameter TNO
has an observed VR-band magnitude of
around~27 (assuming\footnote{\edit1{As shown in \citet{2020A&A...638A..23F}, among other results,
TNOs may have albedos that range from 3\% to 30\% (ignoring the high albedos of the very largest TNOs, which are very unlikely to be present in our sample). The albedo assumption of 10\% used here simply allows us estimate our required depth. In all papers related to this program where we discuss TNO size we will use the appropriate albedo as estimated from dynamical class, and include appropriate uncertainties on size, since we will make no measurements of albedo in this program.}} an albedo of~0.1).
To reach this depth with DECam requires an exposure time of nearly 4~hours; details of this estimate are given in DEEP paper~II (Trujillo et al.). We have therefore designed an observational survey in which each pointing is observed for four hours of nearly-continuous open shutter time.

\subsection{Observing strategy \label{sec:strategy}}

Our complete observing strategy is presented in detail in Trujillo et al.\ (Paper~II); here we present a brief summary.

Our program is divided evenly between the A and B semesters.
In each semester, we identify two patches on the sky that can be observed for four hours each and can be combined into a single whole night; each patch is located near the ecliptic plane.
Our observing seasons are roughly April--June and August--September.
Our patches are named A0, A1 (A semester) and B0, B1 (B semester).
Each patch 
covers a roughly triangular region on the sky,
with a kingpin field (A0a, A1a, B0a, B1a) that anchors the patch.
%(Figure~\ref{fig:chad}).
The size of the patch grows each year to accommodate the
dispersion on the sky of our targets due to differential apparent rates of motion. The slowest moving objects are expected to stay in the kingpin field, whose coordinates are given in Paper~II,
and the rates of the fastest moving objects dictate the number of pointings needed in a patch in each year.
%As shown in Figure~\ref{fig:chad}, 
In Year~1 (2019) each patch has three fields; Year~2 (nominally, 2020) each patch has six fields (the three from Year~1, plus three new ones); and in Year~3 (nominally, 2021) each patch has ten fields (three from Year~1, three from Year~2, and four new ones from Year~3).
\edit1{In other words, the number of fields accumulates in each year: all of the Year~1 fields are also observed in Years~2 and~3; all of the Year~2 fields are also observed in Year~3.}
Each field requires 0.5~night, so the total observing request for these opposition pointings sum to \{6,12,20\}~nights per year.

In 2019 we also observed our Year~1 fields (a--c) in off-opposition pointings 
to provide an additional strong constraint on orbit solutions. These off-opposition pointings required an additional 6~nights of telescope time. 
Some additional time was allocated in 2019 to extend our program, totalling 46.5~nights.

Our survey began in 2019A, and
operated normally in 2019A and 2019B. However, all of our 2020A 
and the first half of our 2020B
observing time was lost to COVID-related shutdown of CTIO.
We were allocated time in 2022A and 2022B (Year~4) and 2023A (Year~5) to make up for lost 2020A and 2020B time. Because our TNOs continue to disperse on the sky, our 2022A, 2022B, and 2023A telescope time --- which is approximately equal to the time lost in 2020, when our TNOs inhabited a more compact distribution on the sky --- is used to sample, rather than cover completely, the region of sky where our 2019-discovered TNOs are located in 2022 and 2023.

%Our science goals require detecting KBOs with diameters as small as 50~km.

%This request was approved and allocated by NOAO (with a small amount of extra time allocated in 2019A; future semesters have not been scheduled yet).
%However, in Year 1 of our program each field will be imaged in both VR and g filters, which
%enables a color measurement for each object, so that our total time request is {6,6,10} nights in
%Years {1,2,3} in each of the A and B semesters, for a total request of 44 nights over six semesters.

%\begin{figure}
%\plottwo{decamplot.png}{astrometricuncertainty.png}
%\caption{DEEP survey simulation that was used to plan our observing program.
%{\em Left:} 
%Black (Years~1--3), blue (Years~2--3), and red
%(Year~3) fields and simulated detections are shown; grey
%shows non-detections.
%Recovery efficiency is as shown.
%The full density simulation is not shown here,
%for clarity.
%The ``kingpin'' fields are in the lower right of the 10-pointing pattern.
%{\em Right:}
%Astrometric uncertainty after our survey is completed for different simulated classes of TNOs, with gaussian noise of 0.6~arcseconds included. Solid lines mark the median orbit uncertainty while dashed lines bound the 90\% confidence region for each orbit type.
%}
%\label{fig:chad}
%\end{figure}

%\subsection{Expected yield}

Nominally, our three year program and increasing patch size would imply that many of our TNOs would have two year orbital arcs. We call these ``gold standard'' objects, and this population will enable studies of physical properties as a function of dynamical class.
Additionally, some objects will appear in Year~2 data that were not within the patch footprint in Year~1 --- for example, a fast moving object that moves into the A0a kingpin field from the west. These objects may have a one year arc, and are referred to as ``silver standard'' objects. Finally, in Year~3 some objects may move into the field that have never been seen before; we call these
``bronze standard'' objects.

Two years after the conclusion of our survey, almost all of our ``gold standard'' objects will have positional uncertainties on the sky of less than
10~arcseconds,
%(Figure~\ref{fig:chad}),
which makes them feasible targets for JWST spectroscopy (given the field
of view of the acquisition cameras/modes). Even 10~years after our survey’s completion, nearly all
of these objects will have an uncertainty that is still less than the width of a single DECam chip
(540~arcsec), so individual objects would still be readily recoverable.

\edit1{As described in detail in Paper II (Trujillo et al.), we choose 120~second exposures in all cases. This provides a balance between carrying out a maximally efficient survey with minimal overheads, and producing data that can be useful to study main belt asteroids with minimal trailing losses.
All our exposures employ sidereal tracking, with the non-sidereal TNO motions detected through digital tracking, as described below.}

\subsection{Survey simulator}

%\emph{Pedro: adjust language as needed to be in line with rest of paper}
We have developed a survey simulation software package to allow both our team and the community to fully characterize detection efficiency, enabling model comparisons to our sample of objects both in the single night and multi-year regimes. Our efficiency characterization spans a wide range of physically reasonable parameters in orbital elements and photometric properties (magnitudes and light curves), so that both our detected objects and also non-detections are statistically meaningful. 
This work will be presented in detail in Paper~III (Bernardinelli et al.) and will be made available in a Github repository.
This repository will also include a number of example scripts and Jupyter Notebooks demonstrating our  science results, ensuring our results can be reproduced by the wider outer Solar System community. 

\subsection{Linking across nights, runs, and seasons}

The science goals of our project require linking objects across nights, runs, and seasons to produce relatively high fidelity orbit solutions. Linking also provides a high-confidence method for confirming marginal detections from a single night.
Briefly, the inferred motion of any candidate can be used to verify that the object is re-detected at the expected position and velocity in observations either in subsequent nights or runs.
Our observing strategy provides one month arcs in Year~1, and then recoveries in Year~2 and again in Year~3. At the conclusion of this project, our {\em gold standard} objects will have two-year arcs and well-known orbits.
Initial results from our linking are presented
in Paper~VI (Smotherman et al.).
Because our cadence was modified by the 2020 COVID-related shutdown and extended into Years~4 and~5, there will be a small number of objects with 3~or 4~year arcs.

\subsection{Synthetic objects}%: MH and CF}
\label{s:task:synth}

It is difficult, if not impossible, to anticipate the correlations among and cumulative effects from various survey characteristics (e.g., orbits, apparent magnitudes, colors, light curves, stellar crowding, detector sensitivity, chip gaps, seeing, atmospheric transparency, etc).
It is far easier to realistically insert synthetic detections derived from a model population directly into individual exposures.  These synthetic sources can be used to determine detection efficiency and to
develop, test, and compare algorithmic approaches for finding faint objects in our data stream \citep{lawler}.
%(2) verify that what is found matches or deviates from what is expected; 
%(3) check where the search routines are sensitive.

%By having a calibrated synthetic population that experiences the same processing as real sources we can characterize the detection biases induced by our method. Our findings will be calibrated to account for such observational effects, revealing the true underlying population.
We
characterize the detection biases induced by our method using a calibrated synthetic
population. These synthetic sources undergo the same processing as real sources
providing an accurate means for comparing the resulting detection efficiency from the
synthetic population with the actual detections. This comparison yields an effective
representation of the true underlying population.

We use a PSF (point spread function) model to insert synthetic detections into copies of the calibrated exposures at the appropriate RA/Dec using each image's WCS (world coordinate system) solution (to get to pixel-position).
The exposures with implanted synthetic detections are processed through our digital tracking and search algorithms described. We compare the \emph{full} synthetic population to the \emph{fraction recovered} to determine the rate of recovery of synthetic objects within the stacked detections from a single night.
Because the control population is generated from a set of orbits, we can also characterize the night-to-night and season-to-season linking efficiencies.  
Having the single-night detection efficiency for a wide range of observational properties will allow us to explore the biases affecting any object discovered that is inconsistent with one of our modelled populations.   Furthermore, by using the same model population to insert detections in any recovery exposures, even on other observing platforms, we can account for follow-up biases. 

This blind search will establish the detection efficiency for our survey with respect to the observable properties of Solar System objects, including; magnitude limits, dependence on rate and direction-of-motion and light curve amplitude/period, linking efficiency, etc. 
Characterizing our survey will allow us to determine the underlying unbiased population distribution.

\section{Data processing}

\subsection{Pre-processing}

Our data are processed through the DECam Community Pipeline \citep{valdes2014}.
The data are calibrated photometrically and astrometrically. The resulting InstCal images are retrieved from the NOIRLab data archive. 
%xxx DG/UM -- anything to say here? xxx

\subsection{Detection of TNOs in individual images}

%xxx DG xxx
The typical 5-sigma single-exposure depth for a two-minute DEEP field exposure is approximately VR$\sim$23.5.
Across the full survey, we estimate $\sim$150 TNOs (and many asteroids, as discussed below) brighter than this detection limit within our DEEP images. The first step in identifying these objects is the generation of transient source catalogs, using a simple stationary-source rejection from nightly co-added images. A streak detection is then performed using a progressive probabilistic straight-line Hough transform \citep{probhough} to identify objects moving linearly across the field. Several vetting steps are then performed on this list of candidate objects, including checks for linearity in Right Ascension, Declination, and time, rate-of-motion cutoffs to discriminate between TNOs and asteroids, and a visual vetting step for our $\sim$50 most likely candidates to verify that the software detection is real. Once the visual vetting is comp lete, we finish with a list of confirmed TNOs detected within the single exposures, for which we have high-cadence astrometry and photometry over a 4-hour long stare. Secondary source extractions and visual identifications of short-stare images are then performed, which ultimately extend the astrometric arcs for these bright TNOs from 4 hours to $>$48~hours. This data processing and results from the single-exposure detections will be presented in detail in DEEP paper IV (Strauss et al.).

\subsection{Faint TNO discovery using digital tracking}

To detect ultra-faint TNOs,
ideally one would track not at the sidereal rate but at a rate that keeps the target object stationary.
However, there are many such objects in a DECam field, and their rates and directions of motion
vary, and are not known {\em a priori}.
Furthermore, because
TNOs move at around 3\arcsec/hour, exposure times longer than it takes for a TNO to cross one resolution element (5--10 minutes) do not result in enhanced SNR because of trailing losses. 

The collected observations consist of series of $\sim$100 two-minute exposures taken over a 4~hour period. (The intervening DECam readout time is around 20~seconds).
Traditional approaches to detecting TNOs rely on the identification of sources within individual images and then linking these sources to generate orbits. This precludes discovery of objects below a single-image detection threshold, greatly constraining the depth to which faint moving objects could be identified in the DEEP dataset.

Alternative techniques --- known as ``shift and stack'' or ``digital tracking'' methods --- have been developed to search for moving sources below the detection limit of any individual image 
\citep{gladman1997,allen2001,B04,holman2004,heinze2015}.
These are fundamentally different from traditional linking in that they assume a trajectory for a moving object and align a set of individual images along that trajectory in order to look for evidence for a source. As (by definition) the orbits of hitherto unknown asteroids are not known, the space of plausible orbits must be exhaustively searched to guarantee discovery to the co-added depth of the dataset. This is conceptually equivalent to creating $O(10^{10})$ coadds, performing detection in each, and retaining the sources above the threshold. Implemented naively, this is a highly computationally intensive task.
Our team has developed and implemented two complementary digital tracking approaches. These will be presented in the
forthcoming Paper V (Napier et al.).

\subsection{Linking digital tracking objects}

The final step in producing our DEEP catalog is linking single-night digital tracking detections across nights, lunations, and years. The digital tracking detections have both information about the object's position and velocity, so both need to be used for an accurate orbit fitting. The linking procedure, however, is more limited by the density of the candidate TNOs
\citep{bernardinelli2020,bernardinelli2022},
and with a data set such as ours, where the majority of the detections correspond to TNOs, treating our detections as a single a sky coordinate is sufficient for an efficient linking procedure. In Paper~VI (Smotherman et al.) we present objects linked with as few as two detections per season, with further detections recovered across multiple years of data. Our orbits use a specialized fitting technique that is aware of the full uncertainties of the digital tracking procedure, leading to precise dynamical classifications of almost all recovered TNOs. 

\section{Demonstration results}

\begin{figure}
%\plotone{2003QL91.ps}
\includegraphics[angle=90,width=6.5in]{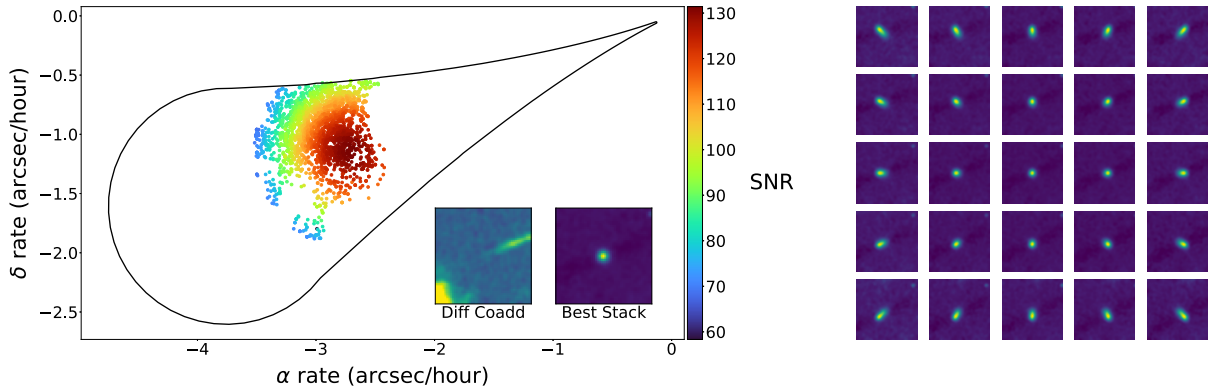}
    \caption{Recovery of the KBO 2003 QL$_{91}$. The panel on the left shows the object's SNR as a function of rate of motion. The black teardrop-shaped region bounds the space of possible rates of motion for bound objects beyond 30 au. Two parameters describing a body's proper motion are sufficient for the detection of KBOs over the 4-hour arcs, during which any nonlinear sky motion \edit1{is} undetectable at the resolution of our images. The panel on the right shows the characteristic elongation of a source as it is stacked at rates deviating from its true rate of motion.}
\label{fig:KNfigure}
\end{figure}

\begin{figure}
%\plotone{2002PC171_245753_cropped.ps}
\includegraphics[angle=90,width=6.5in]{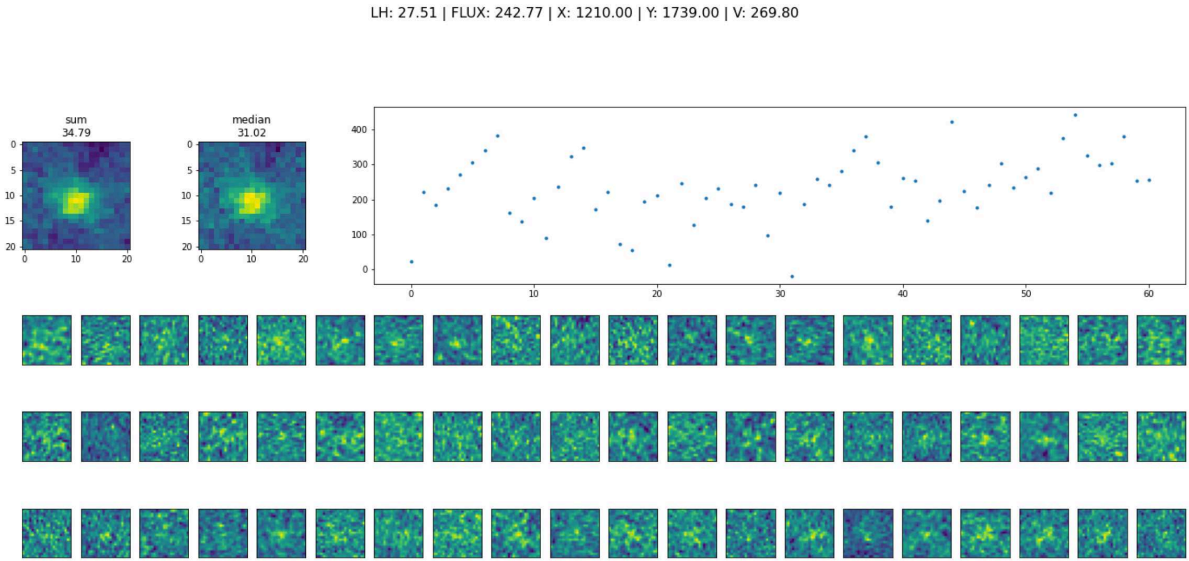}
    \caption{A digital tracking recovery of 2002~PC171, detected in our B1f field on 2021-09-03. This object has $m_V=24.1$. These images are generated using the outputs of \texttt{KBMOD} and are used for human vetting. (All candidates are vetted by multiple people.) The two top left stamps show a simple sum and median coadd of the individual exposures, stacked along the \texttt{KBMOD} trajectory. The top right plot shows the approximate \texttt{KBMOD}-estimated flux in each individual exposure. The bottom three rows show a subset of the individual exposures used in the \texttt{KBMOD} search, aligned along the moving object trajectory. Although one row of stamps was excluded from this figure to save space, all individual stamps are included in the images used for human vetting. Many more details of our digital tracking approach are presented in Papers V (Napier et al.) and VI (Smotherman at al.).}
    \label{fig:KBMOD_2002PC171}
\end{figure}

Here we demonstrate the detection of two TNOs using our team's two independent and complementary detection approaches. Figure~\ref{fig:KNfigure}
shows the recovery of previously known TNO 2003~QL91, a cold classical object with H$\sim$7,
a=43~au, e=0.015, and i=1.5~deg. This object was detected in our digital tracking with SNR of~130 and a VR magnitude of 
around~23.3. This object was detected using 
a pipeline developed at the University of Michigan (see Paper~V: Napier et al.).
In
Figure~\ref{fig:KBMOD_2002PC171}
we show a detection of 2002~PC171
made with KBMOD (the Kernel Based Moving Object Detection), a pipeline developed at University of Washington
\citep{Whidden_2019,Smotherman_2021}.
This is also a classical TNO
(H$\sim$7.6, a=44.8~au, e=0.06, i=3.6~deg), with an
apparent magnitude of around~24.1 that was detected at around SNR=30--35.
These objects are quite bright compared to most TNOs that DEEP will observe, but the overall approach is the same. Of course, the vast majority of DEEP-observed objects will be new discoveries, not previously known objects like these two, but these first results, for known TNOs, helps validate our approaches.
These new detections
are consistent with our expected detection limit near VR=27, and in the following section we describe our expected yield from the entire survey.
Many more details of our digital tracking approach and results are given in Paper V (Napier et al.).

%\subsection{Trans-Neptunian object detections}

% \begin{figure}
% \plotone{2013-gt136.png}
% %    \includegraphics[trim=1in 3.5in 1in 1in, clip, width=\textwidth]{2013_GT136_Sample.png}
%     \caption{\label{fig:KBMOD_gt136} A high-SNR shift-and-stack recovery of TNO 2013 GT136. Top left: the high-SNR co-added image. Remaining panels: a sample of per-epoch cutouts. This was the only {\em known} TNO present in the analyzed field.}
% \end{figure}

\section{Expected survey results \label{sec:expected}}%: DET}
%\MJP{Lots of ``xxx'' in the paragraph(s) below...}

\subsection{Expected sensitivity and yield}

A summary of the expected yield from this program is shown in Table~\ref{tab:yield}.
We use the size distribution of \citet{fuentes2010} to estimate the number of TNOs that we will detect. 
{\em Gold standard} objects will have two or three year arcs. {\em Silver standard} objects will have one year arcs. {\em Bronze standard} objects will have four hour arcs. These three arc lengths are a natural result of the observing strategy described in Section~\ref{sec:strategy} and \edit1{Paper II}.
We expect to observe
thousands of objects with which our science cases can be addressed. With these estimates, our most sensitive investigation --- measuring the size distribution of faint TNOs using only four hour arcs --- should yield
18,000~or more objects. We predict at least 10,000~objects with one year arcs, and at least 5000~objects with two year (or more) arcs, which should allow for robust dynamical classifications.
%xxx put more ETC details here? xxx

\begin{table}
\centering
\begin{tabular}{|c|c|c|c|c|c|c|} \hline
Science & SNR & Exposure & Sensitivity & Gold & Silver & Bronze \\
goal & requirement & time (min) & (VR mags) &  &  &  \\ \hline
{\em Optimistic} \\ \hline
Size dist. & 3 & 240 & 27.5 & 7200 & 14400 & 24000 \\
%Colors & 10 & 240 & 25.5 & 1080 & 2160 & 3600 \\
Lightcurves & 10 & 2 & 24.5 & 360 & 720 & 1200 \\
\hline
{\em Conservative} \\ \hline 
Size dist. & 3 & 240 & 27.0 & 5400 & 10800 & 18000 \\
%Colors & 10 & 240 & 25.0 & 720 & 1440 & 2400 \\
Lightcurves & 10 & 2 & 24.0 & 180 & 360 & 600 \\ \hline
\end{tabular}
\caption{Expected yields from this program for three different TNO science cases. 
{\em Gold standard} objects are those with two (or more) year arcs. {\em Silver standard} objects have one year arcs. {\em Bronze standard} objects have four hour arcs. These three populations naturally result from our observing strategy as described in Section~\ref{sec:strategy} and \edit1{Paper II}.
The two panels show (top/optimistic) the results estimated from the DECam ETC and (bottom/conservative)
a conservative assumption that
assumes 0.5~mag worse than the top panel.
This ideal case has been somewhat disrupted by COVID-related shutdowns at CTIO (in addition to time lost to weather and technical issues), and we will report a final accounting of objects in each category in a future paper.
%Our current data processing results imply a mean 50\% detection magnitude of around~26.8, which is slightly worse than the conservative predictions shown here.
%The top panel shows the depth (and consequent yield) estimated from the DECam Exposure Time Calculator. The bottom panel shows a conservative assumption that the delivered performance is 0.5~mag worse than the top panel, with a commensurate decrease in yield.
%The results presented in Section~\ref{s:task:proc} are in between these two end-cases, though closer to the ETC/optimistic values.
%The approximate diameters for these limiting magnitudes for objects at 
%40~au (for example) are 40~km for the size distribution;
%100~km for colors; and 150~km for lightcurves.
%The limiting magnitude for TNO colors is set by the sensitivity in i~band, since we require detections in both VR and i to measure colors
}
\label{tab:yield}
\end{table}

\subsection{Comparison to existing catalogs}

At present, %220901
in the Minor Planet Center list of TNOs, Centaurs, and Scattered Disk Objects 
there are around 4500~objects 
with preliminary designations
(arc lengths of 24~hours or greater), and slightly more than 1000~objects (with secure, multi-year orbits).
Thus, we will increase the number of objects with preliminary designations --- objects used to determine the size distribution of TNOs --- by a factor 
of 2--3: our
10,800--14,400
{\em silver standard} objects. 
Our 
5400--7200
{\em gold standard} objects will have well-defined orbits, so that each of the above measurements --- size distribution, lightcurves --- can be carried out as a function of dynamical class, yielding deeper understanding of the origin and evolution of the outer Solar System.
Advanced dynamical classification may be possible 
for some DEEP objects, as discussed in a forthcoming paper.
%
%The number of TNOs with measured colors at present is in the hundreds. In this program, because of the potential confusion in correcting individual TNO lightcurves across epochs, we may be limited in our ability to determine the VR-i color for specific objects, but we will be able to measure the average ensemble color very precisely, with several thousand objects contributing. xxx did i talk about color anywhere? xxx
Finally, as of this writing, %220901
there are around 120~TNOs with reliable (U quality code of 2- or better) lightcurves in the Lightcurve Database
\citep{lcdb};
this program will produce hundreds to thousands of partial TNO lightcurves.

\subsection{Comparison to LSST and other surveys}

LSST (the 
Legacy Survey of Space and Time, with first light and the beginning of science operations both in 2024) will survey most of the southern sky in their standard wide, fast, deep cadence, which has a single 30-second image exposure depth of 24.5 (5$\sigma$)
\citep{ivezic2019}.
Thus, the DEEP survey is 2--2.5~magnitudes deeper than the anticipated nominal LSST catalog, which
in turn is around 1--2~magnitudes deeper than our single-exposure DEEP depth.
Furthermore, because the LSST observational cadence is very sparse (a handful of visits over two weeks), applying digital tracking techniques to reach our DEEP depth would require combining frames over years of LSST observing --- an extremely computationally intensive challenge. Thus, the data collected and objects detected here will be, in general, too faint to be detected by LSST. Thus, the DEEP catalog will probe a size regime that LSST may not be able to access.
An exception may be the LSST Deep Drilling fields \citep{ivezic2019}, which have the potential to reach mag$\sim$27, depending on the configuration of the exposures and visits. However, the Deep Drilling pointings that have been established to date are not on the ecliptic and hence are not likely to result in a large number of TNO discoveries.

\begin{figure}
\plotone{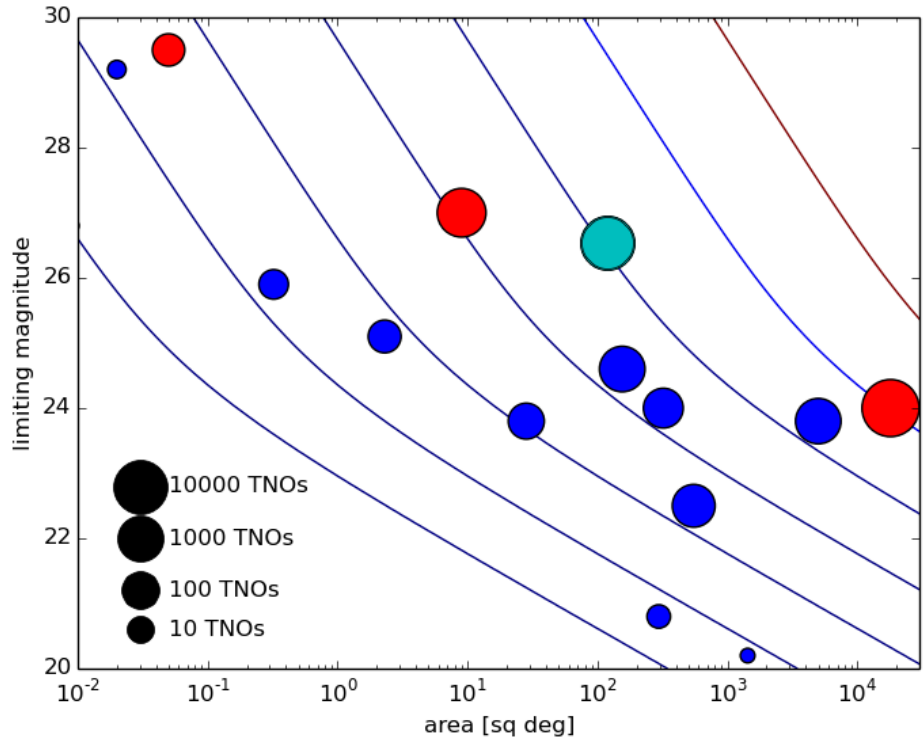}
    \caption{Visualization of survey comparisons among published (dark blue), forthcoming (red), and DEEP (cyan). The forthcoming surveys are LSST wide fast deep, LSST Deep Drilling, and the JWST TNO pencil beam project. The contours (from~1 in the lower left to $10^6$ in the upper right, logarithmically spaced) show the expected number of TNOs to be discovered, assuming ecliptic pointings, using the best fit brightness distribution from B04. The symbol size shows the actual number of objects detected. The Dark Energy Survey, for example, which is shown by the symbol 
    (5000, 23.8)
``underperforms'' relative to expectations because the majority of the sky coverage is away from the ecliptic.
    Our DEEP survey will produce the largest catalog of TNOs, and of faint TNOs, to date.}
        \label{fig:surveys}
\end{figure}

Figure~\ref{fig:surveys} shows a comparison of the DEEP depth and coverage to other TNO surveys. The DEEP combination of depth, through digital tracking, and wide area enabled by DECam is unprecedented in TNO science. The manifestation of this will be in our very large high fidelity catalog and the new science that will be produced, by our team and by others. The only ground-based survey that will be deeper than DEEP is the relatively small-scale LSST Deep Drilling Field program; the other, fainter surveys are from space-based platforms (HST, JWST).

\subsection{Main Belt Asteroids}

While studying main belt asteroids (MBAs)
is only a secondary goal of this project, 
some cadence decisions were made to optimize MBA science, as described in Paper II (Trujillo et al.), and this program will produce a very powerful
and large-scale sample of asteroids at relatively faint magnitudes. 
Our observing cadence includes ``short stares'' at each field on two nights when the field is not visited in a long stare, which will produce 48~hour arcs and therefore decent orbits for most main belt asteroids.
We estimate approximately 500--1000 asteroids per field 
down to the single frame
limiting magnitude
of VR$\sim$22.5--24.5~\edit1{mag}. MBAs move sufficiently fast (around 30~arcsec/hour) that 
each visit to any pointing presents
an entirely new set of asteroids in the field.
Therefore, 
across our entire survey, we expect to detect around 50,000~unique MBAs. 
For each of these asteroids we will measure (partial) lightcurves, and the ensemble of all targets can be used to derive mean asteroid shape as a function of size, semi-major axis, etc.
We will also identify asteroids with unusual rotational lightcurves
\edit1{(for example, objects with short rotation periods that exhibit multiple clear periods within our four-hour sequences)}.
Detailed results will be presented in a future paper.

\subsection{Extensions of this project}

It is likely that our survey will reveal a number of objects that appear intrinsically interesting, perhaps through having unusual orbits, lightcurves, or other properties. 
If additional observations are needed that fall outside the scope of the DEEP program, our team 
will carry out follow-up observations as appropriate.

%\vspace{2ex}
%
%\noindent Finally, we note that it has been proposed to locate at least one LSST Deep Drilling Field on the ecliptic plane. If that plan is implemented, one strategically interesting decision could be placing that LSST DDF location on or adjacent to our DEEP survey patches. This would allow xxx

A requirement of our NOIRLab Survey program is to deliver catalogs and ancillary products to
the NOIRLab archive and data services (in addition to the images, which will also be publicly available). Thus, additional science can be carried out with either our moving object catalogs or the sidereal catalogs and time-series photometry that 
will be a natural byproduct of our data processing. The description of this data delivery will be discussed in a future paper.

\section{Forthcoming papers}

Our team will produce a large number of 
DEEP science papers over the coming years. Our first set
consists of six papers, including this one. The other papers are the following:

\begin{itemize}
    \item {\bf Paper II.} Trujillo et al.\ describe our survey strategy in detail.
    \item {\bf Paper III.} Bernardinelli et al.\ present our survey simulator.
    \item {\bf Paper IV.} Strauss et al.\
describe science results from
TNOs detected at single-exposure depth.
    \item {\bf Paper V.} Napier et al.\ 
present digital tracking results for single epoch detections.
    \item {\bf Paper VI.} Smotherman et al.\
describe TNOs that are linked across multiple visits.
\end{itemize}

Future papers in the series include but are not limited to TNO dynamical studies; asteroid detections in DEEP data; and a wide range of papers mining the DEEP catalog to address the outstanding science questions presented in this paper.

\section{Conclusions}

We are carrying out a 46.5~night NOIRLab (formerly NOAO) survey entitled DEEP: The DECam Ecliptic Exploration Project. The main goals are to detect thousands of previously unknown TNOs to magnitude~VR$\sim$27 (diameters $\sim$50~km) in order to address a range of science questions related to the formation and evolution of the Solar System.
We will detect thousands of TNOs --- perhaps as many as 18,000~with four hour arcs (or better) --- thus increasing the 
catalog of known TNOs by a factor of~4 and
transforming the catalog of known TNOs and our understanding of the outer Solar System.
This paper is the first in a series of papers that present technical approaches and science results. 

\acknowledgments

This work is based in part on observations at Cerro Tololo Inter-American Observatory at NSF’s NOIRLab (NOIRLab Prop. ID 2019A-0337; PI: D. Trilling), which is managed by the Association of Universities for Research in Astronomy (AURA) under a cooperative agreement with the National Science Foundation.
We acknowledge the terrific support from NOIRLab staff for this program. 

This work is supported by the National Aeronautics and Space Administration under grant No.\ NNX17AF21G issued through the SSO Planetary Astronomy Program and by the National Science Foundation under grants No.\ AST-2009096 and AST-1409547. This research was supported in part through computational resources and services provided by Advanced Research Computing at the University of Michigan, Ann Arbor. This work used the Extreme Science and Engineering Discovery Environment \citep[XSEDE; ][]{XSEDE}, which is supported by National Science Foundation grant number ACI-1548562. This work used the XSEDE Bridges GPU and Bridges-2 GPU-AI at the  Pittsburgh Supercomputing Center through allocation TG-AST200009.

H. Smotherman acknowledges support by NASA under grant No.\ 80NSSC21K1528 (FINESST). H. Smotherman, M. Juri\'{c} and P. Bernardinelli acknowledge the support from the University of Washington College of Arts and Sciences, Department of Astronomy, and the DiRAC Institute. The DiRAC Institute is supported through generous gifts from the Charles and Lisa Simonyi Fund for Arts and Sciences and the Washington Research Foundation. M.\ Juri\'{c} wishes to acknowledge the support of the Washington Research Foundation Data Science Term Chair fund, and the University of Washington Provost’s Initiative in Data-Intensive Discovery. 

This project used data obtained with the Dark Energy Camera (DECam), which was constructed by the Dark Energy Survey (DES) collaboration. Funding for the DES Projects has been provided by the US Department of Energy, the US National Science Foundation, the Ministry of Science and Education of Spain, the Science and Technology Facilities Council of the United Kingdom, the Higher Education Funding Council for England, the National Center for Supercomputing Applications at the University of Illinois at Urbana-Champaign, the Kavli Institute for Cosmological Physics at the University of Chicago, Center for Cosmology and Astro-Particle Physics at the Ohio State University, the Mitchell Institute for Fundamental Physics and Astronomy at Texas A\&M University, Financiadora de Estudos e Projetos, Funda\c{c}\~{a}o Carlos Chagas Filho de Amparo \'{a} Pesquisa do Estado do Rio de Janeiro, Conselho Nacional de Desenvolvimento Cient\'{i}fico e Tecnol\'{o}gico and the Minist\'{e}rio da Ci\^{e}ncia, Tecnologia e Inova\c{c}\~{a}o, the Deutsche Forschungsgemeinschaft and the Collaborating Institutions in the Dark Energy Survey.

The Collaborating Institutions are Argonne National Laboratory, the University of California at Santa Cruz, the University of Cambridge, Centro de Investigaciones En\'{e}rgeticas, Medioambientales y Tecnol\'{o}gicas–Madrid, the University of Chicago, University College London, the DES-Brazil Consortium, the University of Edinburgh, the Eidgen\"{o}ssische Technische Hochschule (ETH) Z\"{u}rich, Fermi National Accelerator Laboratory, the University of Illinois at Urbana-Champaign, the Institut de Ci\`{e}ncies de l’Espai (IEEC/CSIC), the Institut de Física d’Altes Energies, Lawrence Berkeley National Laboratory, the Ludwig-Maximilians Universit\"{a}t M\"{u}nchen and the associated Excellence Cluster Universe, the University of Michigan, NSF’s NOIRLab, the University of Nottingham, the Ohio State University, the OzDES Membership Consortium, the University of Pennsylvania, the University of Portsmouth, SLAC National Accelerator Laboratory, Stanford University, the University of Sussex, and Texas A\&M University.


\vspace{5mm}
\facility{Blanco (DECam)}


\software{}




\bibliography{deep_paper1_revised}{}
\bibliographystyle{aasjournal}



\end{document}